\documentclass[journal]{IEEEtran}
\usepackage[T1]{fontenc}
\usepackage[utf8]{inputenc}
\usepackage{authblk}
\usepackage{mathrsfs}
\usepackage{pifont}
\usepackage{bbding}
\usepackage{tipa}
\usepackage{color}
\usepackage{cite}
\usepackage{epsfig}
\usepackage{graphicx}
\usepackage{url}
\usepackage{amsfonts}
\usepackage{multicol}
\usepackage{float}
\usepackage{times}
\usepackage{psfrag}
\usepackage{subfigure}
\usepackage{stfloats}
\usepackage{amsmath}
\usepackage{footnote}
\usepackage{array}
\usepackage{amsmath,epsfig}
\usepackage{stmaryrd}
\usepackage{amssymb}
\usepackage{epic}
\usepackage{graphicx}
\usepackage{curves}
\usepackage{balance}
\usepackage[linesnumbered,ruled,vlined]{algorithm2e}
\usepackage{algpseudocode}
\usepackage{multirow}
\usepackage{amsmath}
\usepackage{xcolor}
\usepackage{graphics}
\usepackage{epsfig}
\usepackage{epstopdf}
\usepackage{enumerate}
\usepackage{color}
\usepackage{lettrine}
\usepackage{graphicx}
\usepackage{threeparttable}
\usepackage{bm}
\usepackage{amsmath,epsfig,amssymb,subfigure,bm,dsfont}
\usepackage{lipsum}
\usepackage{stfloats}
\usepackage{booktabs}

\begin{document}
\bstctlcite{IEEEexample:BSTcontrol}
\newtheorem{theorem}{Theorem}
\newtheorem{proposition}{Proposition}
\newtheorem{definition}{Definition}
\newtheorem{lemma}{Lemma}
\newtheorem{corollary}{Corollary}
\newtheorem{remark}{Remark}
\newtheorem{construction}{Construction}
\newtheorem{problem}{Problem}
\newtheorem{alg}{Algorithm}[section]

\newcommand{\supp}{\mathop{\rm supp}}
\newcommand{\sinc}{\mathop{\rm sinc}}
\newcommand{\spann}{\mathop{\rm span}}
\newcommand{\essinf}{\mathop{\rm ess\,inf}}
\newcommand{\esssup}{\mathop{\rm ess\,sup}}
\newcommand{\Lip}{\rm Lip}
\newcommand{\sign}{\mathop{\rm sign}}
\newcommand{\osc}{\mathop{\rm osc}}
\newcommand{\R}{{\mathbb{R}}}
\newcommand{\Z}{{\mathbb{Z}}}
\newcommand{\C}{{\mathbb{C}}}
\newcommand*{\affaddr}[1]{#1} 
\newcommand*{\affmark}[1][*]{\textsuperscript{#1}}
\newcommand*{\email}[1]{\texttt{#1}}
\newcounter{TempEqCnt}
\newcounter{mytempeqncnt}

\title{DNN-aided Read-voltage Threshold Optimization for MLC Flash Memory with Finite Block Length}

\author{{Cheng Wang, Kang Wei, Lingjun Kong, Long Shi, Zhen Mei, Jun Li, and Kui Cai}
\thanks{
C. Wang, K. Wei, and J. Li are with School of Electronic and Optical Engineering, Nanjing University of Science and Technology, Nanjing, P. R. China (e-mail:\{cheng.wang, kang.wei, jun.li\}@njust.edu.cn).
L. Kong is with College of Telecommunication and Information Engineering, Nanjing University of Posts and Telecommunications, Nanjing, P. R. China (e-mail: ljkong@njupt.edu.cn)
L. Shi, Z. Mei, and K. Cai are with Science and Math Cluster, Singapore University of Technology and Design, Singapore (e-mail: slong1007@gmail.com, mei\_zhen@outlook.com, cai\_kui@sutd.edu.sg).
.
}}

\maketitle
\begin{abstract}
The error correcting performance of multi-level-cell (MLC) NAND flash memory is closely related to the block length of error correcting codes (ECCs) and log-likelihood-ratios (LLRs) of the read-voltage thresholds.
Driven by this issue, this paper optimizes the read-voltage thresholds for MLC flash memory to improve the decoding performance of ECCs with finite block length.
First, through the analysis of channel coding rate (CCR) and decoding error probability under finite block length, we formulate the optimization problem of read-voltage thresholds to minimize the maximum decoding error probability.
Second, we develop a cross iterative search (CIS) algorithm to optimize read-voltage thresholds under the perfect knowledge of flash memory channel.
However, it is challenging to analytically characterize the voltage distribution under the effect of data retention noise (DRN), since the data retention time (DRT) is hard to be recorded for flash memory in reality.
To address this problem, we develop a deep neural network (DNN) aided optimization strategy to optimize the read-voltage thresholds, where a multi-layer perception (MLP) network is employed to learn the relationship between voltage distribution and read-voltage thresholds.
Simulation results show that, compared with the existing schemes, the proposed DNN-aided read-voltage threshold optimization strategy with a well-designed LDPC code can not only improve the program-and-erase (PE) endurance but also reduce the read latency.
\end{abstract}
\section{Introduction}\label{Sec:introduction}
\IEEEPARstart{N}{AND} flash memory is widely used over the past decade due to low power consumption and large storage capacity.
The original NAND flash memory cell can only store one bit with two levels, which is called single-level-cell (SLC).
Using the multi-level-cell (MLC) or triple-level cell (TLC) technique~\cite{kim2008future, cai2012error, lee2017ldpc}, the flash memory can store multiple bits over a single memory cell.
However, as the number of levels in each memory cell increases, serious scaling challenges loom up in the NAND flash memory, resulting in a negative effect on the reliability.
These challenges originate from the characteristics of flash devices that can be seen as several noise models, such as programming noise (PN), cell-to-cell interference (CCI), random telegraph noise (RTN), and data retention noise (DRN)~\cite{li2014noise}.

Among various noises in flash memory, the DRN is caused by the charge leakage at the floating-gate of flash memory cells~\cite{Cai2015Data}.
The charge leakage starts when a flash memory cell is programmed.
The overall period of this process is called the data retention time (DRT).
As the size of memory chip decreases, the floating-gate of a flash memory cell stores much fewer electrons, which degrades the performance of flash memory.
This is due to the fact that a small amount of charge leakage has remarkable influence on the floating-gate transistor.
Compared with SLC, the MLC technology intensifies the decoding errors caused by the DRN, as the reduced interval of write voltage at each storage state distorts the voltage distribution of flash memory.
As a result, the increasing number of program-and-erase (PE) cycles and the DRT limit the operational lifetime of flash memory.

To improve the reliability of flash memory, hard-decision error correcting codes (ECCs), such as Bose-Chaudhri-Hocquenghem (BCH) and Reed-Solomon (RS) codes were employed in flash memory~\cite{cho2014block,Chen2008Error}.
To enhance the decoding error performance of ECCs,~\cite{Dong2011On, Wang2014Enhanced, chen2018rate, Wang2019Adaptive} proposed the utilization of soft decision in flash memory.
Later on, various soft-decision decoding algorithms were proposed to achieve desirable error correcting performance.
For example, the belief-propagation (BP) algorithm is one of the probability-based iterative decoding algorithms with excellent performance~\cite{gallager1962low, xiao2004graph, sharon2007efficient,Wei2019Page}.
It is well known that LDPC codes are decoded with soft information such as channel log-likelihood-ratios (LLRs).
In order to achieve better error-correcting performance, the soft-decision decoder demands more reliable and accurate soft information that can be obtained by the read process~\cite{Dong2011On, Wang2014Enhanced, Aslam2016Read,Aslam2014Non,Mei2019On}.
For the flash memory channel, the problem of obtaining soft information can be turned into that of optimizing the read-voltage thresholds~\cite{Aslam2016Read}.

Driven by this observation, much effort has been put into the optimization of read-voltage thresholds~\cite{Dong2011On,Wang2014Enhanced,Aslam2016Read,Peleato2015Adaptive,Wei2018Read}.
The well-designed read-voltage thresholds can convert hard information (i.e., voltages of cells) into soft information (i.e., LLRs), which greatly improve the decoding performance of flash memory.
Initially, flash memory employed the hard-decision memory sensing that utilizes the hard information generated by the fixed read-voltage thresholds.
However, the hard-decision method is only effective when the flash memory noise is small.
To prolong the lifetime of flash memory, the soft read-voltage sensing strategy becomes a prevailing solution for flash memory.
Prior works in~\cite{Dong2011On, Aslam2016Read} introduced a nonuniform memory sensing strategy to reduce the memory sensing precision and read latency while maintaining good error-correction performance.
These works obtain the read-voltage thresholds by utilizing entropy value of each unreliable region.
Nevertheless, the optimization of read-voltage thresholds relies on extensive simulations and the memory sensing level is limited.
To solve this dilemma, the work in \cite{Wang2014Enhanced} developed an adjustable sensing strategy for multiple reads of the same flash memory cell, which selects the word-line voltages by maximizing the mutual information (MMI) between the input and output of the equivalent discrete read channel.

However, the existing works have the following issues.
First, the aforementioned threshold optimization strategies did not take into account the block length of ECCs that used in flash memory~\cite{Dong2011On, Aslam2016Read, Wang2014Enhanced}.
Notably, the block length of ECCs for emerging memories are usually short due to stringent requirements on low decoding complexity and read latency.
In practice, there is an significant gap between the actual channel coding rate (CCR) and capacity of the flash memory model in~\cite{Aslam2016Read} under finite block length~\cite{Polyanskiy2010Channel}.
Recent research has unveiled that the flash memory channel after sensing by read-voltage thresholds can be regarded as a discrete memoryless channel (DMC)~\cite{Wang2014Enhanced}.
Several theoretical approaches investigated the threshold optimization in DMC from the perspective of information theory~\cite{kurkoski2014quantization,romero2016ldpc}.
Following these theoretical approaches, we characterize the maximum coding rate in flash memory as a function of block length and error probability.
Building upon the rate analysis, we optimize the read-voltage thresholds for flash memory.

Second, the prior works in~\cite{Dong2011On,Aslam2016Read,Wang2014Enhanced} designed the read-voltage thresholds for flash memory assuming perfect knowledge of PE cycles and DRT.
In practice, it is rather difficult to record DRT.
Without the knowledge of PE cycles and DRT, the following methods were proposed to recover the soft information of flash memory channel under the effect of the DRN.
A flash correct-and-refresh technique proposed in~\cite{Cai2012Flash} read the data stored in flash memory periodically and utilized the ECCs to perform the decoding and reprogramme the flash memory.
Later on, \cite{Lee2014Decision} developed a decision-directed estimation (DDE) algorithm to remit the DRN by utilizing a Gaussian mixture model to estimate the voltage distribution of flash memory.
The DDE algorithm first compares the input and output of the decoder to find the best-fit parameters of the Gaussian model, and then utilizes the Gaussian model to adjust the read-voltage thresholds.
Recently, a retention-aware belief-propagation (RABP) decoding scheme was proposed to combat the DRN in MLC flash memory~\cite{aslam2017retention}.
If the decoding fails, the RABP algorithm adjusts the input LLRs based on the decoded bits and performs another round of decoding.
Furthermore,~\cite{aslam2018decision} proposed a RABP aided channel update algorithm to estimate the voltage distribution of MLC flash memory.
It regards voltage distribution of flash memory as Gaussian distribution and utilizes the decoding results to update the mean and variance of voltage distribution.
However, the decoding processes in~\cite{Cai2012Flash,Lee2014Decision,aslam2017retention,aslam2018decision} result in either large energy consumption or high decoding latency, which contradicts with practical use of flash memory.
In addition, these methods are applicable only when the DRN is within a small certain range such that the decoder can still provide sufficient correct information.
In this context, these methods cannot handle the errors caused by the DRN that exceeds the correction capability of ECCs.

Recently, rapid development of deep learning inspires us to handle the variation of flash memory channel caused by the DRN.
With an explosive increase in big data, the deep learning technologies, such as deep neural network (DNN), can distill the data effectively and extract abstract correlations from data~\cite{D2018Deep,Liang2018An}.
For the flash memory, in contrast to the existing methods that require a round of decoding to obtain the useful information, the DNN allows the system to train a model off-line and explore the relationship between the input and output, and the well-trained DNN model can directly generate the information from the processed data.
These findings motivate us to design a DNN-aided read-voltage optimization strategy that does not rely on the knowledge of DRT.
%




The primary goal of this paper is to optimize the read-voltage thresholds in MLC flash memory with finite ECC block length.
Towards this goal, we first formulate the optimization problem of read-voltage thresholds under finite block length, and then propose the cross iterative searching (CIS) algorithm and DNN-aided optimization strategy to optimize the read-voltage thresholds, respectively.
The main contributions of this paper are summarized as follows:
\begin{itemize}
  \item \emph{Read-voltage threshold optimization under finite block length---}We study the CCR of MLC flash memory under finite block length and optimize the read-voltage thresholds with perfect knowledge of PE cycles and DRT.
        Under finite block length, we first formulate the read-voltage optimization problem to maximize the CCR by minimizing the maximum error probability.
        Then, we develop a CIS algorithm to solve this problem.
        Simulation results show that, compared with MMI-based quantization and entropy-based quantization, the proposed CIS algorithm can significantly improve the lifetime of flash memory.

  \item \emph{DNN-aided read-voltage threshold optimization---}We develop a DNN-aided optimization strategy to optimize the read-voltage thresholds without the knowledge of DRT.
      The core of the proposed DNN-aided scheme is to train a multi-layer perception (MLP) network to learn the relationship between the voltage distribution (i.e., input of the MLP) and the read-voltage thresholds (i.e., output of the MLP).
        Simulation results show that, compared with the RBAP decoding scheme, the DNN-aided scheme can not only improve the PE endurance but also reduce the read latency.
\end{itemize}

The remainder of this paper is organized as follows.
Section~\uppercase\expandafter{\romannumeral2} presents the MLC flash memory channel model and investigates its CCR under finite block length.
Section~\uppercase\expandafter{\romannumeral3} formulates the read-voltage thresholds optimization problem under finite block length and proposes the CIS algorithm.
Section~\uppercase\expandafter{\romannumeral4} proposes the DNN-aided optimization strategy.
Section~\uppercase\expandafter{\romannumeral5} shows the simulation results.
Section~\uppercase\expandafter{\romannumeral6} concludes this paper.
\section{System Model}
\subsection{Channel Model of MLC NAND Flash Memory}\label{flash memory model}
Let $\mathcal{S}=\{s_0,s_1,s_2,s_3\}$ denote the storage states of MLC flash memory.
A flash memory cell must be erased before programming.
Let $s_{0}$ denote the erased state of an MLC flash memory cell.
With the reference to~\cite{Aslam2016Read}, the voltage distribution of the cell at state $s_{0}$ is approximately modeled as a Gaussian distribution $p_{\text{e}}(v)=\mathcal{N}(\mu_{\text{e}},\sigma_{\text{e}}^2)$ with mean $\mu_{\text{e}}$ and standard deviation $\sigma_{\text{e}}$, respectively.
In addition, let $s_1$, $s_2$, and $s_3$ denote the programmed states. 
Moreover, the voltages at these programmed states are generated by using an incremental step-pulse programming technique.
Then, the voltage distribution of the cell at each programmed state follows a uniform distribution~\cite{Kang1995A}:
\begin{equation}\label{eq:ISPP}
p_{\text{p}_{s_i}}(v)=
\begin{cases}
1/V_\text{p}, & v\in \left[V_{s_i} V_\text{p}\right)\\
0, & \text{elsewhere,}
\end{cases}
 \quad  \; i=1,2,3,
\end{equation}
where $V_\text{p}$ denotes the programming step voltage and $V_{s_i}$ denotes the target programmed voltage of $s_i$.

The MLC flash memory channel is generally attenuated by the PN, cell-to-cell interference (CCI), RTN and DRN~\cite{Aslam2016Read,dong2012estimating,dong2010using}.
\subsubsection{Programming Noise}
Let $n_\text{pn}$ denote the PN. The voltage programming process is influenced by the PN, which can be approximately modeled as a Gaussian distribution $n_\text{pn}(v)=\mathcal{N}(0,\sigma_{\text{pn}}^2)$ with zero mean and standard deviation $\sigma_\text{pn}$~\cite{Takeuchi1996A}.
The programming process does not change the voltage of erased state, but only effects the voltage distributions of states $s_1$, $s_2$, $s_3$~\cite{Aslam2016Read}.


\subsubsection{Cell-to-cell Interference}
Let $n_\text{c}$ denote the CCI.
As the major noise source in the MLC flash memory~\cite{Cai2017Error,Aslam2016Read,dong2012estimating}, the CCI results from the scaling down of the flash memory chip, leading to a voltage shift $V_\text{C}$ among the cells:
\begin{equation}\label{CCI}
{V_\text{C}} = \sum\limits_j {\Delta {V_j}{\zeta _j}}, 
\end{equation}
where $\Delta V_j$ represents the voltage variation of the $j$-th interfering cell programmed after the victim cell, and $\zeta_j$ represents the coupling coefficient between the $j$-th interfering cell and the victim cell.
The effect of CCI can be estimated and the pre-distortion/post-compensation technique can be employed to mitigate the influence of CCI~\cite{dong2010using}.
However, this technique cannot eliminate the CCI's effect on the erased state $s_{0}$. 
Let $V_{s_\text{0}}$ denote the target voltage of the erased state.
According to~\cite{Aslam2016Read, dong2010using}, the voltage distribution of the cell for even-bit line and odd-bit line at the erased state is modeled by two Gaussian distributions, i.e., $n_c^{\text{even}}=\mathcal{N}(\tilde \mu_\text{e}^{\text{even}},\sigma_{\text{e}}^2)$ and $n_c^{\text{odd}}=\mathcal{N}(\tilde \mu_\text{e}^{\text{odd}},\sigma_{\text{e}}^2)$, with the same variance $\sigma_{\text{e}}^2$ and different means:
\begin{subequations}
\begin{equation}
\tilde \mu_\text{e}^{\text{even}} = V_{s_\text{0}} + {V_{\text{mean}}}(2{K_\text{x}} + {K_\text{y}} + 2{K_{\text{xy}}}),
\end{equation}
\begin{equation}
\tilde \mu_\text{e}^{\text{odd}} = V_{s_\text{0}} + {V_{\text{mean}}}({K_\text{y}} + {K_{\text{xy}}}),
\end{equation}
\end{subequations}
where ${V_{\text{mean}}}=(V_{s_\text{0}}+V_{s_\text{3}})/2-V_{s_\text{0}}$; $\tilde \mu_\text{e}^{\text{even}}$ and $\tilde \mu_\text{e}^{\text{odd}}$ represent the variances of voltage for the even-bit line and odd-bit line cells, respectively; $K_\text{x}$, $K_\text{y}$, and $K_{\text{xy}}$ are the coupling coefficients of the floating gate in the horizontal, vertical, and diagonal directions, respectively.
\begin{figure}
 \centering
    \includegraphics[width=3.2in,angle=0]{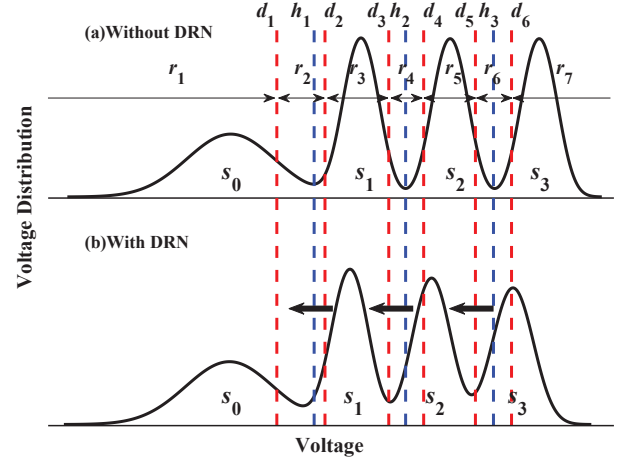}
    \caption{Voltage distribution and 6-level read quantization of an MLC flash memory.}
     \label{fig:v_d}
\end{figure}
\subsubsection{Random Telegraph Noise}
Let $n_\text{rtn}$ denote the RTN. The RTN can be approximately modeled as a Gaussian distribution $n_\text{rtn}(v)=\mathcal{N}(0,\sigma_{\text{rtn}}^2)$ with zero mean and standard deviation $\sigma_\text{rtn}$, where $\sigma_\text{rtn}$ varies with the number of program-and-erase (PE) cycles in a power-law form~\cite{Aslam2016Read}.
From~\cite{aslam2018decision}, $\sigma_\text{rtn}=0.00027(N_\text{PE})^{0.64}$ with $N_\text{PE}$ being the number of PE cycles.

Fig. 1 (a) illustrates the voltage distribution of an MLC flash memory cell under the effect of PN, CCI, and RTN.
\subsubsection{Data Retention Noise}
Let $n_\text{d}$ denote the DRN. The DRN is approximated as a Gaussian distribution $n_{\text{d}_i}(v)=\mathcal{N}(\mu_{\text{r}_{s_i}},\sigma_{\text{r}_{s_i}}^2)$, $i=0$,1,2,3, where $\mu_{\text{r}_{s_i}}$ and $\sigma_{\text{r}_{s_i}}$ are the data-dependent mean and standard deviation, respectively~\cite{Aslam2016Read,dong2010using}. Both
$\mu_{\text{r}_{s_i}}$ and $\sigma_{\text{r}_{s_i}}$ are time-varying and voltage-dependent:
\begin{subequations}
\begin{equation}
\mu_{\text{r}_{s_i}}=\text{log}(1+T)(V_i-V_{\text{0}})[\beta_0(N_\text{PE})^{\alpha_0}+\beta_1(N_\text{PE})^{\alpha_1}],
\end{equation}
\begin{equation}
\sigma_{\text{r}_{s_i}}=0.4\left|\mu_{\text{r}_{s_i}}\right|,
\end{equation}
\end{subequations}
where $T$ is the DRT, $\alpha_0$, $\alpha_1$, $\beta_0$, and $\beta_1$ are constants.

Finally, the overall voltage distribution functions, calculated by the convolution integral of initial voltage distribution functions with various noise functions~\cite{aslam2018decision}, are given by
\begin{gather}\label{equ:pdf}
\begin{aligned}
&p_{s_i}(v)=\frac{1}{{\sigma_{s_i}}\sqrt{2\pi}}e^{-\frac{(v-\mu_{s_i})^{2}}{2\sigma_{s_i}}},i=0,1,2,3,\\
\end{aligned}
\end{gather}
where
\begin{subequations}\label{equ:PN}
\begin{equation}
\mu_{s_{\text{0}}}=V_{s_{\text{0}}}-\mu_{\text{r}_{s_{\text{0}}}},
\end{equation}
\begin{equation}
\sigma_{s_{\text{0}}}=\sqrt{\sigma_{\text{e}}^{2}+\sigma_{\text{rtn}}^{2}+\sigma_{\text{r}_{s_{\text{0}}}}^{2}},
\end{equation}
\begin{equation}
\mu_{s_{\hat i}}=V_{s_{\hat i}}-V_\text{p}/2-\mu_{s_{\hat i}},
\end{equation}
\begin{equation}
\sigma_{s_{\hat i}}=\sqrt{\sigma_{\text{pn}}^{2}+\sigma_{\text{rtn}}^{2}+\sigma_{\text{r}_{s_{\hat i}}}^2},\hat i=1,2,3.
\end{equation}
\end{subequations}


According to~\cite{aslam2018decision}, the parameters of MLC flash memory are set as $V_{s_{0}}=1.4$, $V_{s_{1}}=2.6$, $V_{s_{1}}=3.2$, $V_{s_{3}}=3.93$, $V_\text{p}=0.2$, $\sigma_{\text{e}}=0.34$, $\sigma_{\text{pn}}=0.05$, $\beta_{\text{0}}=0.00001$, $\beta_{\text{1}}=0.00008$, $\alpha_0=0.68$, and $\alpha_1=0.52$, respectively.
From \eqref{equ:pdf}, the increase of either $N_\text{PE}$ or DRT changes the voltage distribution, which causes the read errors and degrades the endurance of flash memory.

\subsection{Read-voltage Quantization}
\begin{figure}
 \centering
    \includegraphics[width=3.2in,angle=0]{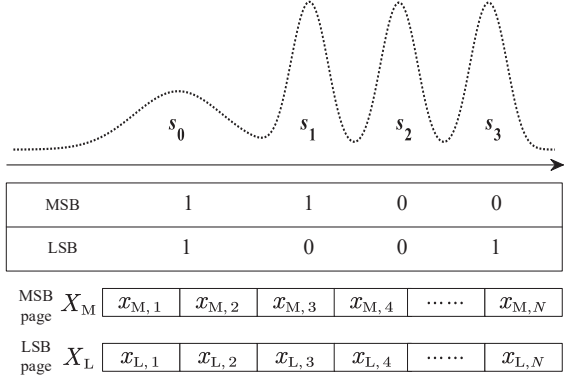}
    \caption{Coding for MLC flash memory.}
     \label{fig:page}
\end{figure}
For the MLC flash memory, the relationship among the block, cell wordline/bitline, and page is briefed as follows~\cite{Cai2017Error}.
Each memory block contains multiple rows of cells.
Each cell stores $K=2$ bits, i.e., the most significant bit (MSB) and least significant bit (LSB).
To reduce the raw bit error rate, the Gray coding is used to map the 2 bits in each cell to one of the storage states.
As shown in Fig.~\ref{fig:page}, the storage states $s_{0},s_{1},s_{2},s_{3}$ correspond to the information bits 11, 10, 00, 01, respectively.
The MSBs of all cells on the same wordline are combined to form an MSB page, and the LSBs of all cells on the same wordline are combined to form an LSB page.

ECC is used to detect and correct the raw bit errors that occur within flash memory.
In this paper, we use two independent length-\emph{N} ECC to encode the input sequence of the MSB and LSB pages as $X_\text{M}=(x_{\text{M},1},x_{\text{M},2},...x_{\text{M},N})$ and $X_\text{L}=(x_{\text{L},1},x_{\text{L},2},...x_{\text{L},N})$, respectively.
During the write process in the \emph{n}-th cell, every $K=2$ bits, i.e., $(x_{\text{M},n},x_{\text{L},n})$ are first mapped to a storage state.
Then, according to the storage state of a memory cell, the programming operation shifts the voltage of this cell to a well-designed write-voltage threshold.
During the read process, to transform the voltage value into soft information (i.e., LLRs) for ECC decoding, the readback voltages need to be quantized by comparing with precomputed read thresholds.

Consider a voltage quantization strategy with \emph{J}-level reads.
The read voltages of memory cells are quantized into \emph{J+1} regions.
Let $\mathcal{D}=\{d_{1},d_{2},\ldots,d_{J}\}$ collect \emph{J}-level read-voltage thresholds, and $\mathcal{R}=\{r_{0},r_{1},\ldots,r_{J}\}$ collect \emph{J+1} output regions where $r_j=[d_j,d_{j+1}]$ with $d_{0}=0$ and $d_{J+1}=+\infty$.
In addition, the read-voltage thresholds of flash memory cells yield $0<d_{1}<d_{2}<\cdots<d_{J}$.
Fig.~\ref{fig:v_d} illustrates this quantization with 6-level read.
For $j=1,2,\cdots,J$ and $k=1,2,\cdots,K$, the initial LLR of the $k$-th bit in the $j$-th region is calculated by
\begin{equation}
{L(j,k)} = \text{log}\frac
{\int\limits_{{d_{j - 1}}}^{{d_j}}\sum\limits_{i\in \mathcal{Q}_k} {{p_{s_i}}(v)\text{d}v}}
{\int\limits_{{d_{j - 1}}}^{{d_j}}\sum\limits_{i=1}^{4} {{p_{s_i}}(v)\text{d}v}-{\int\limits_{{d_{j - 1}}}^{{d_j}}\sum\limits_{i\in \mathcal{Q}_k} {{p_{s_i}}(v)\text{d}v}}},
\label{eq:LLR}
\end{equation}
where $\mathcal{Q}_k$ is the set of states each with the $k$-th bit being 1.
Based on~\eqref{eq:LLR}, we can obtain the LLR of each region.

The choice of read-voltage thresholds determines the LLRs, thus has great impact on the ECC decoding performance.
Therefore, the goal of this paper is to maximize the read reliability of MLC flash memory by optimizing the read-voltage thresholds.


\subsection{CCR for Flash Memory Channel under Finite Block Length}
%

A DMC comprises of an input set, output set, and a probability transition matrix where the probability distribution of the output depends only on
the input at that time and is conditionally independent of previous channel inputs or outputs.
Since the read process transforms storage states into discrete region values, the flash memory channel can be treated as a DMC.

Let $W:\mathcal{S}\rightarrow \mathcal{R}$ denote the DMC with transition probabilities $W(r_j|s_i)$, $s_i\in \mathcal{S}$, $r_j\in \mathcal{R}$, where input $s_i$ and output $r_j$ correspond to the storage state and quantization region, respectively.
The transition probability function of the voltage region $r_j$ given input $s_i$ is
\begin{gather}\label{Conditional Probability}
\begin{align}
W\left( r_j|s_i \right)&=w_{r_{j},s_{i}}=\int_{d_{j}}^{d_{j+1}}{p_{s_{i}}(v)\text{d}v}\nonumber\\
&=Q\left(\frac{d_{j}-\mu_{s_{i}}}{\sigma_{s_{i}}}\right)-Q\left(\frac{d_{j+1}-\mu_{s_{i}}}{\sigma_{s_{i}}}\right),
\end{align}
\end{gather}
where $p_{s_{i}}(v)$ is given in \eqref{equ:pdf} and $Q(\epsilon)=\int_{\epsilon}^{\infty}{\frac{1}{\sqrt{2\pi}}e^{\frac{-t^{2}}{2}}\text{d}t}$.
Moreover, the probability of output $r_j$ is given by
\begin{gather}\label{Output Distribution}
\begin{align}
P(r_j)&=p_{r_{j}}=\sum_{r_{j}\in \mathcal{R}}{p_{s_{i}}w_{r_{j},s_{i}}}\nonumber\\
=&\sum_{r_{j}\in \mathcal{R}}{p_{s_{i}}\left[Q\left(\frac{d_{j}-\mu_{s_{i}}}{\sigma_{s_{i}}}\right)-Q\left(\frac{d_{j+1}-\mu_{s_{i}}}{\sigma_{s_{i}}}\right)\right]}.
\end{align}
\end{gather}
From \eqref{Conditional Probability} and \eqref{Output Distribution}, the mutual information between input $s_i$ and output $r_j$ is
\begin{gather}\label{Mutual Information1}
\begin{align}
I\left( P,W \right) &=\sum_{s_i\in \mathcal{S}}{\sum_{r_j\in \mathcal{R}}{P\left( s_i \right) W\left( r_j|s_i \right) \log}}\frac{W\left( r_j|s_i \right)}{P\left( r_j \right)}
\nonumber\\
&=\sum_{s_i\in \mathcal{S}}{\sum_{r_j\in \mathcal{R}}{p_{s_i}}w_{r_j,s_i}\log \frac{w_{r_j,s_i}}{p_{r_j}}},
\end{align}
\end{gather}
and the unconditional information variance is
\begin{align}\label{Unconditional Information Variance1}
&U\left( P,W \right) =\sum_{s_i\in \mathcal{S}}{\sum_{r_j\in \mathcal{R}}{P\left( s_i \right) W\left( r_j|s_i \right) }}\left(\log\frac{W\left( r_j|s_i \right)}{P\left( r_j \right)}\right)^2\nonumber\\
&=\sum_{s_i\in \mathcal{S}}{\sum_{r_j\in \mathcal{R}}{p_{s_i}}w_{r_j,s_i}\left( \log \frac{w_{r_j,s_i}}{p_{r_j}} \right) ^2}-\left[ I\left( P,W \right) \right] ^2.
\end{align}

As~\cite{Polyanskiy2010Channel} unveiled, for a finite block length code and DMC, the achievable CCR with a given error probability $\epsilon$ and a code block length \emph{N} yields
\begin{gather}\label{Upper Bound}
\begin{aligned}
R(N,\epsilon,\gamma)\geq &I(P,W)-\sqrt{\frac{U(P,W)}{N}}Q^{-1}(\epsilon)+\frac{\log{N}}{2N},\\
\end{aligned}
\end{gather}
where $Q^{-1}$ is the inverse function of $Q(\epsilon)$.
\section{Read-voltage Threshold Optimization for MLC Flash Memory}
In this section, we give the upper bound of decoding error probability for MLC flash memory channel and formulate the read-voltage threshold optimization.
Unlike conventional methods such as MMI and entropy-based quantization, our optimization problem focuses on finite block length.
\subsection{Error Performance under Finite Block Length}
First, we rewrite \eqref{Upper Bound} as
\begin{gather}\label{Upper Bound1}
\begin{align}
Q^{-1}(\epsilon)&\geq\mathcal{T}(N,\epsilon,\gamma, P, W),
\end{align}
\end{gather}
where
\begin{flalign}\label{function_T}
&\mathcal{T}(N,\epsilon,\gamma, P, W)=\nonumber&\\
&\quad\quad\left[I(P,W)-R(N,\epsilon,\gamma)+\frac{\log{N}}{2N}\right]\sqrt{\frac{N}{U(P,W)}}.&
\end{flalign}
For the flash memory, both $I$ and $U$ vary over different $P$ and $W$, since $P$ and $W$ depend on the parameters of flash memory, such as number of PE cycles, DRT and read-voltage thresholds according to \eqref{equ:pdf} and \eqref{Conditional Probability}.
Thus, the function $\mathcal{T}$ in \eqref{function_T} can also be interpreted as a function with respect to these parameters:
\begin{flalign}\label{function T_f}
&\mathcal{T}(N,\bar{R},{\mathcal{D}},E,T)=&\nonumber\\
&\quad\quad\quad\left[I(\mathcal{D},E,T)-\bar{R}+\frac{\log N}{2N}\right]\sqrt{\frac{N}{U(\mathcal{D},E,T)}},&
\end{flalign}
where $\bar{R}$ is the code rate of ECCs used in flash memory.

As $Q$ function is monotonically decreasing, the decoding error probability is upper bounded by $\epsilon \leq$ $Q\left(\mathcal{T}(N,\bar{R},{\mathcal{D}},E,T)\right)$.
Thus the maximum error probability is $\epsilon_{\rm max} = Q\left(\mathcal{T}(N,\bar{R},{\mathcal{D}},E,T)\right)$.
In this context, our goal is to optimize the read-voltage thresholds by minimizing the maximum decoding error probability:
\begin{gather}\label{Optimization Function1}
\begin{aligned}
\mathcal{D}^*=\mathop{\arg\min}_\mathcal{D} {\epsilon_{\rm max}},
\end{aligned}
\end{gather}
where $\mathcal{D}^{*}$ is the set of optimal read-voltage thresholds.




Due to the write process of MLC flash memory, the MSB and LSB have different channel conditions~\cite{Wang2014Enhanced,Sun2016Explointing}.
Consequently, the error probabilities of MSB and LSB vary over different quantization regions.
Taking the 6-level read in Fig.~\ref{fig:v_d} for example, the MSB errors often occur in region $r_4$, and the LSB errors often occur in regions of $r_2$ and $r_6$~\cite{Sun2016Explointing}.
In addition, according to \eqref{Conditional Probability} and \eqref{Output Distribution}, the transition probabilities $W$ of MSB and LSB, denoted by $W_\text{M}$ and $W_\text{L}$, are diverse.
Furthermore, the decoding error probabilities of MSB and LSB are independent, since independent encoding processes are used for these two pages.
In the view of this independence, the average maximum error probability for MLC flash memory over the two pages is given by
\begin{gather}
\begin{aligned}\label{eq:two_page_epsilong}
\epsilon_{\rm max} = \frac{Q\left(\mathcal{T}_\text{M}\right)+Q\left(\mathcal{T}_\text{L}\right)}{2},
\end{aligned}
\end{gather}
where the $\mathcal{T}$ functions of MSB and LSB are denoted by
\begin{subequations}
\begin{flalign}
\mathcal{T}_\text{M}=\left[I(P,W_\text{M})-\bar{R}+\frac{\log N}{2N}\right]\sqrt{\frac{N}{U(P,W_\text{M})}},
\end{flalign}
\begin{flalign}
\mathcal{T}_\text{L}= \left[I(P,W_\text{L})-\bar{R}+\frac{\log N}{2N}\right]\sqrt{\frac{N}{U(P,W_\text{L})}}.
\end{flalign}
\end{subequations}
Overall, we can formulate the optimization problem as
\begin{subequations}\label{power}
\begin{flalign}
\quad{\mathcal P}:\quad&\mathop {\min }\quad \epsilon_{\rm max}\\
&~~\text{s.t.}~\quad 0<d_{1}<d_{2}<\cdots<d_{J}.
\end{flalign}
\end{subequations}


Due to the dimension of $\mathcal{D}$, analytical solution of $\mathcal{P}$ is computationally intractable.
In the following, we develop an efficient method to solve this problem.
%

\subsection{Cross Iterative Searching Algorithm}
In this part, we utilize genetic algorithm and CIS algorithm to optimize the read-voltage thresholds in various read-levels.
In the genetic algorithm, the evolution is implemented by using a set of stochastic genetic operators to mimic the natural process of reproduction and mutation.
Although the genetic algorithm can solve complex problems, high quality solutions require massive computations to explore the entire search space for global optimization~\cite{Ali2007GSR}.
For our problem, the computation of genetic algorithm dramatically increases as the dimension of $\mathcal{D}$ goes larger.
To reduce the complexity, the cross iterative searching algorithm helps us to find local optimum solution within certain region which saves a lot of time.
Combining the genetic algorithm and cross iterative searching algorithm, we can escape from local optimum and obtain near-optimal results.



As shown in \textbf{Algorithm} 1, we develop a CIS algorithm to solve the optimization problem given in (19a).
In the read-voltage threshold optimization, all the read-voltage thresholds are constrained by (19b).
Before the iterative searching process, the CIS algorithm needs to determine the initial value of the read-voltage thresholds (see line 1 of \textbf{Algorithm} 1).
The well-designed initial value will accelerate the convergence speed and avoid trapping into local optimum.
Let $\mathcal{H}=\{h_1, h_2, h_3\}$ denote a set that collects the read-voltage thresholds under hard decision.
We can identify the hard-decision thresholds by letting
\begin{equation}\label{hard_threshold}
\begin{split}
p_{s_0}(v=h_1)=p_{s_1}(v=h_1),\\
p_{s_1}(v=h_2)=p_{s_2}(v=h_2),\\
p_{s_2}(v=h_3)=p_{s_3}(v=h_3).
\end{split}
\end{equation}
Then, we initialize the $J$-level read-voltage thresholds as $\mathcal{D}^0=\{d^{0}_1,d^{0}_2,\ldots,d^{0}_J\}$, where $d^{0}_1=h_1-\delta$, $d^{0}_j=h_1+(j-1)\delta$, for $j=2,\cdots,J-1$, $d^{0}_J=h_3+\delta$, and $\delta=\frac{h_3-h_1}{J-1}$.

Lines 2-8 show the iterative searching process.
First, the ranges of read-voltage thresholds are determined in order to reduce the searching space (see line 5).
During the ($i+1$)-th iteration, we search $d_j^{i+1}$ over $\left[d_j^i-\lambda,d_j^i+\lambda\right]$,
where $\lambda$ is a well-designed constant (e.g., $\lambda=0.2$ in the \text{simulations}).
Second, the thresholds are updated successively, where each read-voltage threshold is optimized while keeping remaining read-voltage thresholds fixed (see line 6).
Finally, the searching algorithm ends and outputs the optimized read-voltage thresholds if $|\epsilon_{\rm max}^{(i)}-\epsilon_{\rm max}^{(i-1)}|<\rho$ or the maximum number of iterations is reached (see lines 2 and 9).
\begin{algorithm}[t]
\caption{CIS Algorithm}
\LinesNumbered
\KwIn{$\epsilon_{\rm max}$,
maximum iterations $I_{\rm max}$, stopping criteria $\rho$, block length $N$, code rate $\bar{R}$.}
\KwOut{the read-voltage thresholds $\mathcal{D}$.}

{Initialization: \noindent $ i\leftarrow 0$, $\mathcal{D}^0$}\;
\While {$ |\epsilon_{\rm max}^{(i)}-\epsilon_{\rm max}^{(i-1)}|> \rho$\,\, \rm {and}\,\, $i < I_{\rm max}$}
{
$i=i + 1$, $j=1$\;
\While {$j \leq J$}
{
Determine the range of $d_{j}^{(i)}$\;
Search for the local optimal $d_{j}^{(i)}$ using $\underset{d_{j}^{(i)}}{\arg\min } Q(\mathcal{T}(d_j^{(i)}, N, \bar{R}, E, T))$\;
$j=j + 1$;
}
Calculate $\epsilon_{\rm max}^{(i)}$\;
}
{Output $\mathcal{D}^{(i)}$}.
\end{algorithm}
\section{DNN-aided Read-voltage Threshold Optimization}

\subsection{Motivation}
Fig. \ref{fig:v_d} (b) illustrates that the original voltage thresholds become outdated, since the voltage distribution is changed under the effect of DRN in MLC flash memory.
Without the precise read-voltage thresholds, we cannot obtain the correct LLRs in~\eqref{eq:LLR} that depend on these thresholds.
Finally, due to the mismatch between new voltage distribution and outdated read-voltage thresholds, the decoder is unable to decode correctly based on the incorrect LLRs.

From~\eqref{equ:pdf}, the voltage distribution mainly depends on the number of PE cycles and DRT. 
The number of PE cycles for the memory block can be recorded in flash memory~\cite{aslam2018decision}.
Nevertheless, we cannot analytically characterize the voltage distribution under the effect of DRN, since the DRT is hard to be recorded.
Hence, it is great challenging for existing technologies to track the voltage distribution under the effect of DRN.
To address this issue, we design a DNN-aided optimization strategy to optimize the read-voltage thresholds.
\begin{figure}[t]
 \centering
    \includegraphics[width=3.4in,angle=0]{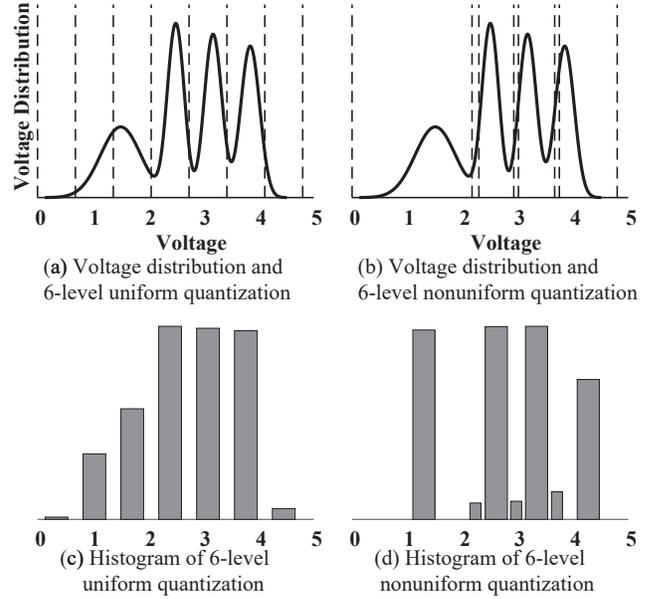}
    \caption{6-level read-voltage quantization for MLC flash memory.}
     \label{fig:quantization_6}
\end{figure}
\subsection{Data Process}
The DNN is a powerful tool to extract deep information from raw data, which can build the non-linear mapping between inputs and outputs~\cite{D2018Deep,Liang2018An}.
However, its learning ability is limited when the input data lacks valuable information. 
For the flash memory, the input data comes from the read process.
Due to the read errors and limited memory sensing precision, it is hard to obtain the accurate voltage of each cell.
In the read process, the read-voltage thresholds can be used to determine the voltage locations over the quantization regions (i.e., the region where each voltage value falls into) and transform each location into a specific LLR of \eqref{eq:LLR}.

In this paper, we adopt the nonuniform quantization to obtain the voltage location information, since the nonuniform read-voltage quantization shows better error-correction performance than uniform under the same number of quantization levels~\cite{Dong2011On,Wang2014Enhanced,Aslam2016Read}.
As an illustration, Fig.~\ref{fig:quantization_6} shows that, under the 6-level quantization, the nonuniform quantization can better capture the characteristics of the voltage distribution, where the histogram is used to count the number of voltage values that fall into each region.
This observation illustrates that the nonuniform quantization can track the variation of voltage distribution under the effect of DRN with limited number of quantization levels.
Therefore, by the nonuniform quantization, the DNN can efficiently learn the relationship between the location information and voltage distribution.

\subsection{Multi-layer Perception Network}
To address the mismatch problem between new voltage distribution and outdated read-voltage thresholds, we propose a DNN-aided decoding strategy to optimize the read-voltage thresholds over different DRT.
Before delving into the proposed scheme, we briefly introduce the DNN.
The MLP is a feedforward DNN which can extract valuable information from extremely complex problems.
In particular, it utilizes a supervised learning technique called backpropagation for training.
A typical MLP network consists of at least three layers and each layer consists of a number of nodes.
The adjacent layers are fully interconnected by weights that are chosen randomly at the beginning.

\begin{figure}[t]
 \centering
    \includegraphics[width=3.4in,angle=0]{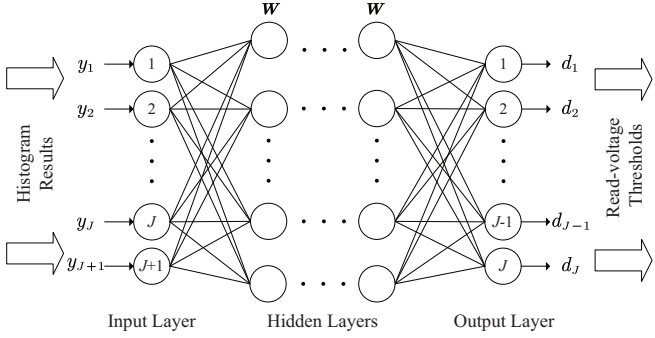}
    \caption{The diagram of an MLP network}
    \label{fig:MLP}
\end{figure}

As shown in Fig.~\ref{fig:MLP}, the MLP is composed of input layer, hidden layers, and output layer.
The input layer that owns $J+1$ nodes receives the input data and forwards it to the hidden layer.
The output layer outputs $\boldsymbol{D}=f\left( \boldsymbol{WY}+\boldsymbol{b} \right) $,
where $\boldsymbol{W}$ and $\boldsymbol{b}$ are the weights and biases of the hidden layer neurons respectively, and $f(\cdot)$ is a non-linear activation function.

For each learning iteration, the MLP receives the input data (i.e., training set, validation set, or test set) and outputs some values.
Based on the error between the MLP output and the expected output (i.e., label), the MLP performs a backpropagation to update the weights of the hidden layers.
By the gradient decent algorithm, the weights are updated by $\boldsymbol{W}(i + 1) = \boldsymbol{W}(i) - \eta  \frac{{\partial E(i)}}{{\partial \boldsymbol{W}(i)}}$,
where $\eta$ is the learning rate and $E(i)$ is the error at $i$-th iteration. With the backpropagation, the DNN can minimize the error between the MLP output and expected output.
\subsection{Training}
\subsubsection{Training Data Generation}
The training data of DNN includes the input data (i.e., histogram results of voltage values) and expected output data (i.e., read-voltage thresholds optimized by \textbf{Algorithm} 1).
As shown in~\eqref{equ:pdf}, the voltage distribution of flash memory channel depends on the number of PE cycles and DRT.
To make the DNN learn the relationship between input data and expected output data, the training set must include the voltage values with different numbers of PE cycles and different DRT.
In addition, the training data is generated within a set of PE cycles $\{4000, 5000, 6000\}$ and a range of DRTs over $[0,10^6]$.
\begin{table}[ht]
 \centering
 \caption{DNN Hyper-Parameters}
 \renewcommand\arraystretch{1.4}
 \setlength{\tabcolsep}{13mm}{
\begin{tabular}{c||c}
\hline
Learning rate& $10^{-5}$\\
\hline
Epoch& 100000\\
\hline
Mini-batch size&500\\
\hline
Initializer&Xavier\\
\hline
Optimizer&Adam\\
\hline
Loss function&MSE\\
\hline
\end{tabular}}
\label{tab2}
\end{table}
\subsubsection{Loss Function}
The loss function is the measurement of errors between the MLP output and expected output.
In our simulations, we employ the mean squared error (MSE) as the loss function, which defined as
\begin{equation}
L_\text{MSE}=\frac{1}{J}\sum\limits_{j = 1}^J {{{({d_j} - {{\hat d}_j})}^2}},
\end{equation}
where $d_j$ and ${\hat d}_j$ are the expected output and MLP output, respectively.

\subsubsection{DNN Parameters}
The sizes of input layer and output layer depend on the read-voltage quantization levels.
In the MLP network, we employ three hidden layers with 512, 256, 128 neurons, respectively.
For each hidden layer and output layer, the activate functions are all \emph{Sigmoid Function}, i.e., $S(x)=\frac{1}{1+e^{-x}}$.
The hyper-parameters are listed in Table~\uppercase\expandafter{\romannumeral1}.

\begin{figure}[t]
 \centering
     \includegraphics[width=3.3in,angle=0]{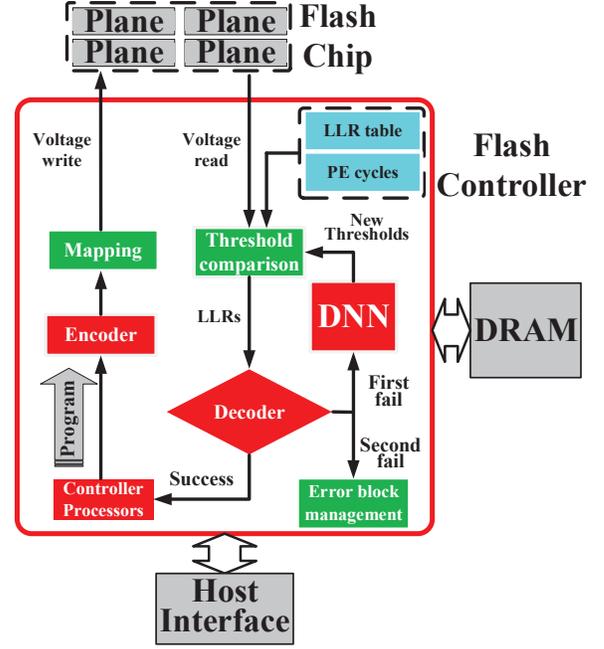}
    \caption{The architecture of DNN-aided MLC flash memory.}
    \label{fig:DNNSensor}
\end{figure}
\subsection{DNN-aided Flash Memory}
In this subsection, we develop a DNN-aided MLC flash memory structure as shown in Fig.~\ref{fig:DNNSensor}.
The DNN is well-trained with the histogram results and the read-voltage thresholds optimized by the proposed CIS algorithm.
First, the controller reads the voltage value from the flash memory chip.
Second, the controller converts these voltage values into LLRs.
Then, the decoder uses these LLRs to perform decoding.
If the decoding fails, the DNN is activated and uses the histogram results to update the read-voltage thresholds.
After that, the decoder receives the updated LLRs and performs decoding again.
If the second decoding fails, the controller records this block as a bad block.

\section{Simulation Results}\label{Sec:numerical}
In the simulations, we use the sum product algorithm as the decoding algorithm where the maximum number of decoding iterations is $I_\text{max}$.
The simulations use three binary LDPC codes, i.e., \emph{2K-QC-code}, \emph{4K-QC-code}, and \emph{2K-random-code}.
In \emph{4K-QC-code}, each entry of a small $7 \times 71$ base matrix $\mathbf{H}_B$ is replaced by either a circulant shift of a $64 \times 64$ identity matrix or a $64 \times 64$ zero matrix.
The block length of this code is 4544 bits and the code rate is set as 0.9.
This irregular code has column-weight of 5 and row-weight of either 50 or 51.
The \emph{2K-QC-code} is chosen as a QC-LDPC code with uniform column-weight of 4 and row-weight of either 40 or 41.
The code rate of \emph{2K-QC-code} are the same as \emph{4K-QC-code}.
The \emph{2K-random-code} is an irregular LDPC code with input and output block length (frame size) of 1998 and 1776 bits, respectively.
The code-rate is 0.89 and the column-weight is 4.

\begin{figure}
\centering
    \includegraphics[width=3.4in,angle=0]{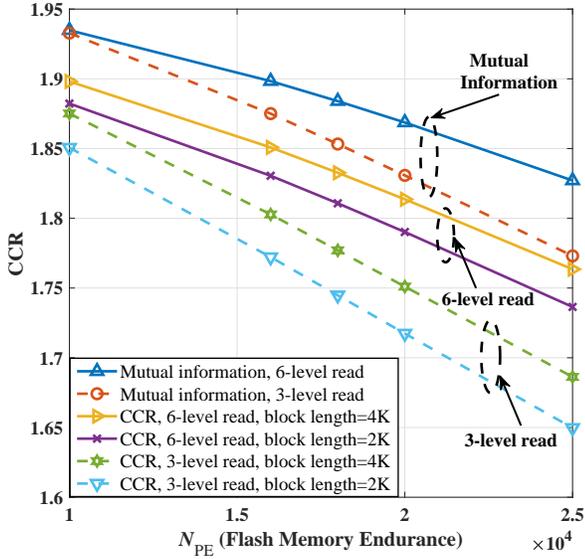}
    \caption{CCR versus PE cycles under different read-level quantization.}
    \label{fig:CCR}
\end{figure}

Fig.~\ref{fig:CCR} plots the CCR under different optimization strategies, read levels, and block length versus PE cycles.
The CCR of mutual information strategy~\cite{Wang2014Enhanced} follows~\eqref{Mutual Information1}, and the CCR of finite block length strategy follows~\eqref{Upper Bound}.
First, it is observed that the loss of CCR enlarges as the block length decreases.
Second, the quantization with larger read-levels contributes to a higher CCR.
This is due to the fact that larger read-level quantization provides more precise voltage information especially with high PE cycles.


\begin{figure}
\centering
    \includegraphics[width=3.4in,angle=0]{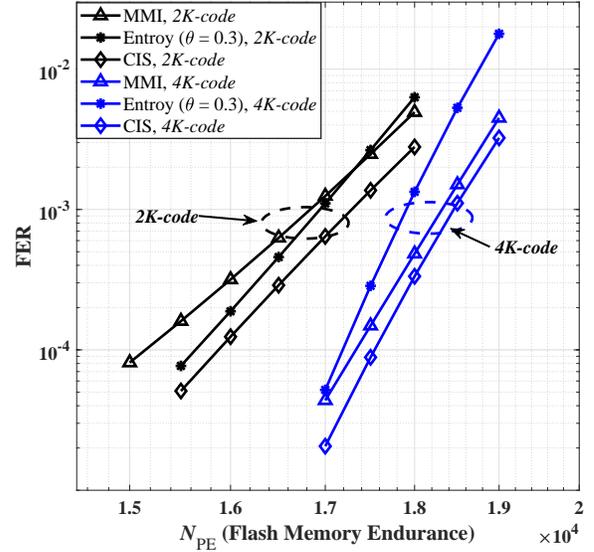}
    \caption{FER performance of LDPC \emph{2K-QC-code} and \emph{4K-QC-code} versus different $N_\text{PE}$ under 6-level read quantization with $I_\text{max}=25$.}
    \label{fig:FER_2K_4K}
\end{figure}
\begin{figure}
\centering
    \includegraphics[width=3.4in,angle=0]{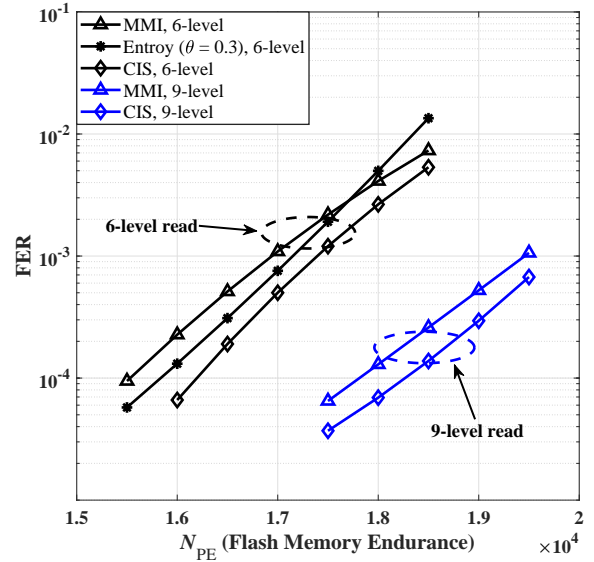}
    \caption{FER performance of LDPC \emph{2K-QC-code} versus different $N_\text{PE}$ under different read-level quantization with $I_\text{max}=30$.}
    \label{fig:FER_2K}
\end{figure}

Fig.~\ref{fig:FER_2K_4K} plots the frame-error-rate (FER) curves over different $N_\text{PE}$ under the proposed CIS algorithm, MMI-based quantization and entropy-based quantization (with the optimized entropy parameter $\theta=0.3$ \cite{Aslam2016Read}) with \emph{2K-QC-code} and \emph{4K-QC-code}, respectively.
Consider that the number of PE cycles ranges over $[15000, 19000]$ and DRT is set to be zero (i.e., $T$ = 0 that represents the early retention time).
It is observed that the proposed CIS algorithm can endure the largest PE cycles among all the three methods.
For example, at FER $=10^{-4}$, the MMI-based quantization and entropy-based quantization with \emph{2K-QC-code} can endure around 15100 and 15600 PE cycles, respectively.
In contrast, the proposed CIS algorithm can extend the endurance limit of PE cycles to 15900.

Fig.~\ref{fig:FER_2K} compares the FER performance versus \text{different} $N_\text{PE}$ between the proposed CIS algorithm, MMI-based quantization, and entropy-based quantization with \emph{2K-QC-code}.
It is observed that the proposed CIS algorithm is superior to both MMI-based quantization and entropy-based quantization under both 6-level and 9-level quantization.
In addition, higher level read quantization performance better.
For example, at FER $=10^{-4}$, the proposed scheme improves the endurance by 2100 PE cycles under the 9-level quantization compared with 6-level read.
This is due to the fact that, with the higher level read quantization, more accurate LLRs are fed into the DDN-aided decoder.
Note that this figure does not show the FER of entropy-based 9-level quantization, since the entropy-based quantization cannot freely choose the read-levels.

\begin{figure}[t]
\centering
    \includegraphics[width=3.4in,angle=0]{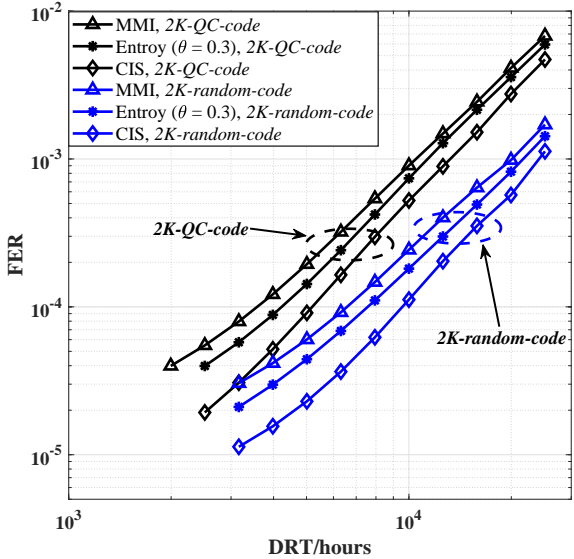}
    \caption{FER performance of LDPC \emph{2K-QC-code} and \emph{2K-random-code} versus different DRT with $N_\text{PE}=8000$ and $I_\text{max}=25$.}
    \label{fig:FER_2K_random}
\end{figure}

Fig.~\ref{fig:FER_2K_random} shows the FER curves versus different DRT between the proposed CIS algorithm, MMI-based quantization, and entropy-based quantization with \emph{2K-QC-code} and \emph{2K-random-code}.
Note that all these quantization methods use the perfect knowledge of DRT and PE cycles.
It is observed that the proposed algorithm is superior to other algorithms with different LDPC codes under the effect of DRN.
For example, at FER $=10^{-4}$, the proposed algorithm can extend the endurance limit of DRT up to 1000 hours and 2000 hours with \emph{2K-QC-code} and \emph{2K-random-code}, respectively.

\begin{figure}[t]
 \centering
    \includegraphics[width=3.4in,angle=0]{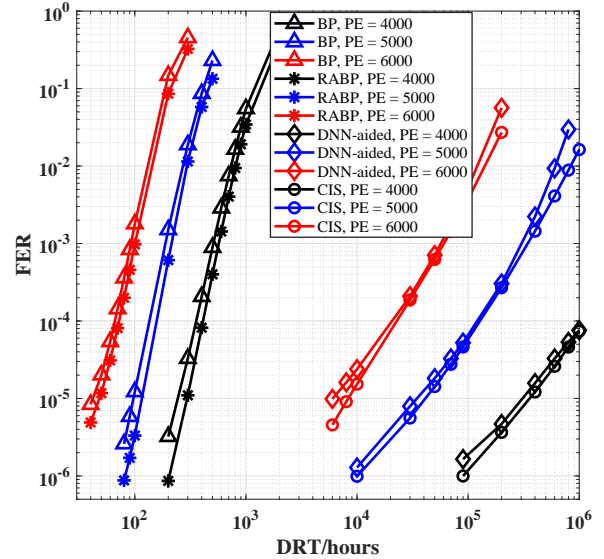}
    \caption{FER performance of LDPC \emph{2K-random-code} under different strategies versus different DRT with $N_\text{PE}=\{4000,5000,6000\}$ and $I_\text{max}=50$.}
    \label{fig:FER_uni}
\end{figure}

Fig.~\ref{fig:FER_uni} plots the FER curves of the BP decoding, the RABP decoding in~\cite{aslam2017retention}, the proposed DNN-aided scheme, and the CIS algorithm.
In this figure, the RABP decoding utilizes the information generated by the first-round BP decoding to amend the LLRs and perform the second-round BP decoding.
First, it is observed that the FER of the proposed DNN scheme approaches that of CIS.
Second, the proposed DNN scheme can significantly improve the tolerance of flash memory against the DRN compared with the BP decoding and RABP decoding.
For example, at FER $=$ $10^{-4}$, the proposed DNN scheme can extend the endurance of flash memory up to nearly 30000, 200000, 1000000 hours, while keeping the $N_\text{PE}$ fixed at 6000, 5000, 4000, respectively.
In addition, the proposed scheme improves the read latency of flash memory compared with the RABP decoding.
This is due to the fact that the RABP decoding demands the second-round decoding to amend the inaccurate results in first-round decoding caused by the DRN.
However, the proposed DNN scheme estimates the read-voltage thresholds every 1000 blocks.
Consequently, there is no need for the proposed scheme to do the second-round decoding, which reduces the read latency.

%
\section{Conclusions}\label{Sec:conclusion}
In this paper, we optimized the read-voltage thresholds for MLC flash memory under finite block length.
First, we analyzed the flash memory channel under finite block length and formulated the threshold optimization problem.
Based on the finite block length theory, we converted the problem of maximizing CCR problem into that of minimizing the maximum decoding error probability.
With perfect knowledge of PE cycles and DRT, we proposed the CIS algorithm to solve this optimization problem.
Furthermore, to address the intractable LLRs under the effect of DRN in reality, we proposed the DNN-aided scheme to optimize the read-voltage thresholds without the knowledge of DRT, where the nonuniform quantization is employed to generate the voltage location information as the input to the MLP.
The simulation results demonstrated that the proposed algorithms improve the PE endurance compared with the existing baseline methods.
In particular, the proposed DNN-aided scheme can reduce the read latency compared with the RABP decoding scheme.


\appendices
\renewcommand{\IEEEQED}{\IEEEQEDopen}

\bibliographystyle{IEEEtran}
\bibliography{reference}

\end{document}